\documentclass{article}
\usepackage{spconf,amsmath,graphicx}
\usepackage{subfig}
\usepackage{algorithmic}
\usepackage{algorithm}
\usepackage{hyperref}
\usepackage{tikz}
\def\checkmark{\tikz\fill[scale=0.4](0,.35) -- (.25,0) -- (1,.7) -- (.25,.15) -- cycle;} 

\title{Classification of the Cervical Vertebrae Maturation (CVM) stages Using the Tripod Network}
\name{\begin{tabular}{c}Salih Atici$^\dagger$$^\ddagger$ \quad Hongyi Pan$^\dagger$$^\ddagger$  \quad Mohammed H. Elnagar* \quad Veerasathpurush Allareddy* \\
Omar Suhaym*$^\star$\quad Rashid Ansari$^\dagger$ \quad Ahmet Enis Cetin$^\dagger$\end{tabular}
\thanks{This work was supported in part by National Science Foundation (NSF) under grant IDEAL 2217023.}
}
\address{$^{\dagger}$Department of Electrical and Computer Engineering, University of Illinois Chicago, Chicago, IL\\
 *Department of Orthodontics,  College of Dentistry, University of Illinois Chicago, Chicago, IL\\
 $^\star$Department of Oral and Maxillofacial Surgery, King Saud bin Abdulaziz \\University for Health Sciences, Riyadh, Saudi Arabia}
\begin{document}

\ninept

\maketitle

\def\thefootnote{$^\ddagger$}\footnotetext{These authors contributed equally to this work.}

%$^\star$ KSAU-HS is an acronym for King Saud bin Abdulaziz University for Health Sciences.}
\def\thefootnote{\arabic{footnote}}
\begin{abstract}
We present a novel deep learning method for fully automated detection and classification of the Cervical Vertebrae Maturation (CVM) stages. The deep convolutional neural network consists of three parallel networks (TriPodNet) independently trained with different initialization parameters. They also have a built-in set of novel directional filters that highlight the Cervical Verte edges in X-ray images. Outputs of the three parallel networks are combined using a fully connected layer. 1018 cephalometric radiographs were labeled, divided by gender, and classified according to the CVM stages.
%The images were cropped to extract the cervical vertebrae using an Aggregate Channel Features (ACF) object detector. 
Resulting images, using different training techniques and patches, were used to train TripodNet together with a set of tunable directional edge enhancers. Data augmentation is implemented to avoid overfitting. TripodNet achieves the state-of-the-art accuracy of 81.18\% in female patients and 75.32\% in male patients. The proposed  TripodNet achieves a higher accuracy in our dataset than the Swin Transformers and the previous network models that we investigated for CVM stage estimation.
\end{abstract}


\section{Introduction}
The success of orthodontic/orthopedic treatment depends on optimal treatment timing, especially in addressing craniofacial skeletal imbalances. The optimal treatment timing relies on identifying craniofacial skeletal maturity stages. Bone age assessment using radiographic analyses was reported to be more accurate than chronological age in determining skeletal maturation, growth rate, the peak period of growth, and the remaining growth potential \cite{ortho1, ortho2, ortho3}. Cervical vertebra maturation (CVM) staging in lateral cephalometric radiographs is a method to determine skeletal maturation. The validity and reliability of the CVM staging have been supported by multiple studies \cite{ortho4, ortho5}. Cervical vertebrae are the first seven bones of the spinal column. Vertebral growth involves changes in the size of vertebral bodies and the shape of the upper and lower borders of C2, C3, C4 vertebrae. These changes have been described into six stages, correlating with morphological modifications of the vertebral shapes. The major limitation of the CVM method is that it is not user-friendly and needs experienced practitioners; researchers reported poor reproducibility among nonexpert examiners \cite{ortho6}. 

The use of machine learning (ML) techniques in the field of medical imaging is rapidly evolving, and a fully automated diagnostic approach has gained attention with its promise of reducing human error as well as the time and effort needed for the task \cite{lee2017deep}. The application of Deep Learning (DL) to study human growth and development from radiographs is a promising idea that needs to be explored. The present study aims to apply a custom-designed DL method to develop a fully automated machine-learning system to detect and classify the CVM stages. There have been recent studies to use pre-trained networks to create fully automated detection and classification of the CVM stages where each utilizes a different dataset \cite{seo2021comparison, dlcvm2}. 
In this study, we propose a custom-designed network model to develop a fully automated system to detect and classify the CVM stages. Our DL network has a tripod-like structure consisting of three parallel networks which are independently trained with different initialization parameters \cite{pan2022multipod}. They also have a built-in set of novel directional filters that highlight the edges of the cervical vertebrae in X-ray images.

\begin{figure*}[htbp]
    \centering
    \includegraphics[width=0.9\linewidth]{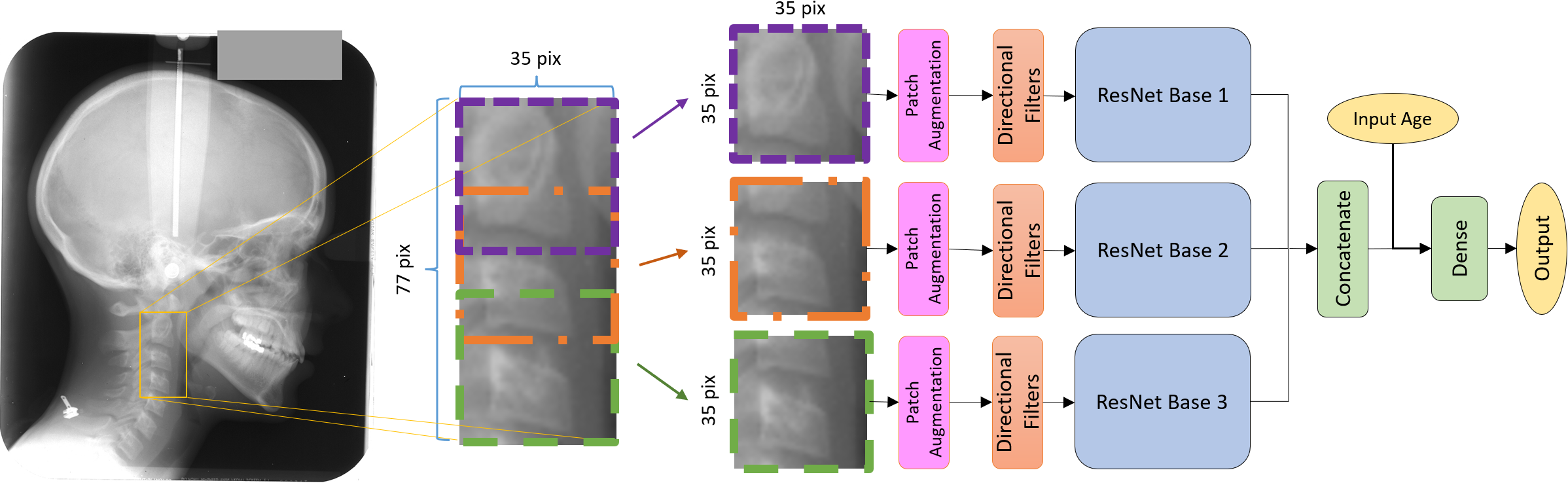}
    \caption{The Model Diagram of the TriPodNet. Overlapping vertebrae images are fed to each pod of the network.}
    \label{fig: model_diag}
\end{figure*}

The TripodNet has some of the features of the transformer networks: (i) Input images are divided into patches, (ii) input patches are augmented as in transformers; and  (iii) the network has a multiheaded structure. ResNet-20 is used as the backbone of TripodNet. Output feature maps of the three parallel ResNets are combined
using a fully connected layer to produce the final decision as shown in Fig.~\ref{fig: model_diag}. Moreover, the age information of the patients is also fed to the network to increase the accuracy of classification. The dataset is divided by gender as male patients can have a different growth rate than female patients.  The proposed model achieves state-of-the-art performance with 81.18\% in female patients and 75.32\% in male patients on the dataset collection, which is superior to the previous results that used DL in a traditional way including the straightforward implementation of the vision transformer \cite{atici2022fully}. 

\section{Methodology}
\subsection{Overview of the CVM Dataset}
\begin{figure}[htbp]
    \centering
    \includegraphics[width=1\linewidth]{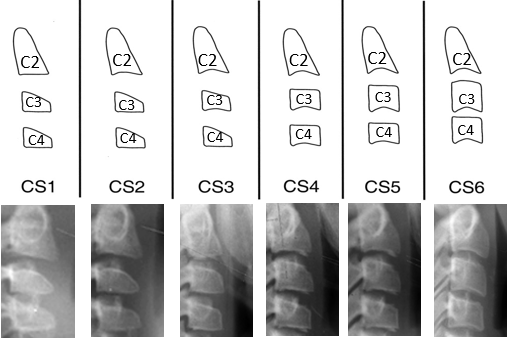}
    \caption{Six CVM Stages and the corresponding X-ray images. The top drawing is adapted from reference~\cite{ortho2}.}
    \label{fig: stages}
\end{figure}

The dataset used in developing our algorithm consists of digitized images of scanned lateral cephalometric films for subjects aged between 4 and 29 obtained from the American Association of Orthodontists Foundation (AAOF) Craniofacial Growth Legacy Collections, an open data source~\cite{AAOF}. The images were studied and labeled by the third author (MHE) who is an expert orthodontist scientist  with more than ten years of experience in classifying CVM. Cervical maturation stages were classified into six stages (CS1- CS6). Six stages are shown in Fig.~\ref{fig: stages} \cite{ortho2}. It is visible from Fig.~\ref{fig: stages} that the main difference among the classes arises in the size and shape of C2, C3, and C4 vertebrae. The change may also happen in one vertebrae only, which may cause confusion for a traditional CNN model which uses the entire image as its input. For example, two classes CS1 and CS2 are only separated by the shape of the lower bound of C2 vertebra whereas C3 and C4 vertebrae have the same shape.

The main dataset was classified into six stages of CVM (CS1-CS6) and we will examine the model performances on the six-stage classification problem. 
The dataset consists of 1012  images classified by the third author (MHE) as the principal evaluator.
Out of 1012 images, 478 images belong to female patients, and 534 images belong to male patients. In our entire dataset, the number of lateral cephalograms belonging to cervical stages CS1, CS2, CS3, CS4, CS5, and CS6 are 153, 182, 174, 159, 167, and 177, respectively. Data augmentation methods such as random translation, AutoContrast~\cite{pytorchautocontract}, AugMix~\cite{hendrycks2019AugMix}, and RandAugment~\cite{cubuk2020randaugment} are implemented with various combinations to avoid the overfitting of the DL networks to the training dataset.

As it is stated earlier, the changes in C2, C3, and C4 vertebrae create 6 classes; therefore, their shape and size play a significant role in the determination of CVM stages. TriPodNet requires three inputs to create an output. In this study, for the network to distinguish the differences among the stages, we used three patches per image where each patch containing one of the C2, C3, or C4 vertebrae. Moreover, every patch is augmented at the beginning of the TriPodNet to increase the performance of each model. Since the size of the dataset is much smaller than the benchmark dataset used to train ResNet-20, data augmentation and patch augmentation methods are necessary to train the TriPodNet. Note that the patch augmentation is different from the data augmentation method we used to avoid overfitting. Rotation and grayscale jittering are implemented to create a different instance of the same input each time an input is about to be fed into the network.

\subsection{Structure of the Multi-Pod Network (MultiPodNet)}

A typical MultiPodNet utilizes two or more parallel Convolutional Neural Networks (CNNs) performing the same computations and may process the input image as sequential image patches in parallel as in transformer networks. The original input image and its augmented versions are fed into convolutional networks forming the MultiPod network and the output feature maps of parallel convolutional networks are concatenated before the fully connected dense layer. In this study, we use TriPodNet because it generated the best results in our dataset and it is the best performer among the other models compared to our arXiv manuscript \cite{pan2022multipod}. Similar to \cite{pan2022multipod}, we also use ResNets as the baseline network in this article. The TriPodNet model  structure is shown in Fig~\ref{fig: model_diag}.

We first segment a given head and neck X-ray image and identify the spine (cervical vertebrae) region using the so-called Aggregate Channel Features (ACF) object detector \cite{acf} as shown in Fig.~\ref{fig: model_diag}. This avoids the process of manually cropping the spine region in each image in the database. As a result, the skull, jaw, and irrelevant background regions are removed before the images are applied to the deep learning algorithm. The ACF object detector automatically extracts the Region of Interest (RoI) in the images thereby reducing the search space of the deep learning structure. Because all the segmented images have variable sizes, they are resized to a common size of $77\times35$. 

\begin{figure}[htbp]
    \centering
    \includegraphics[width=1\linewidth]{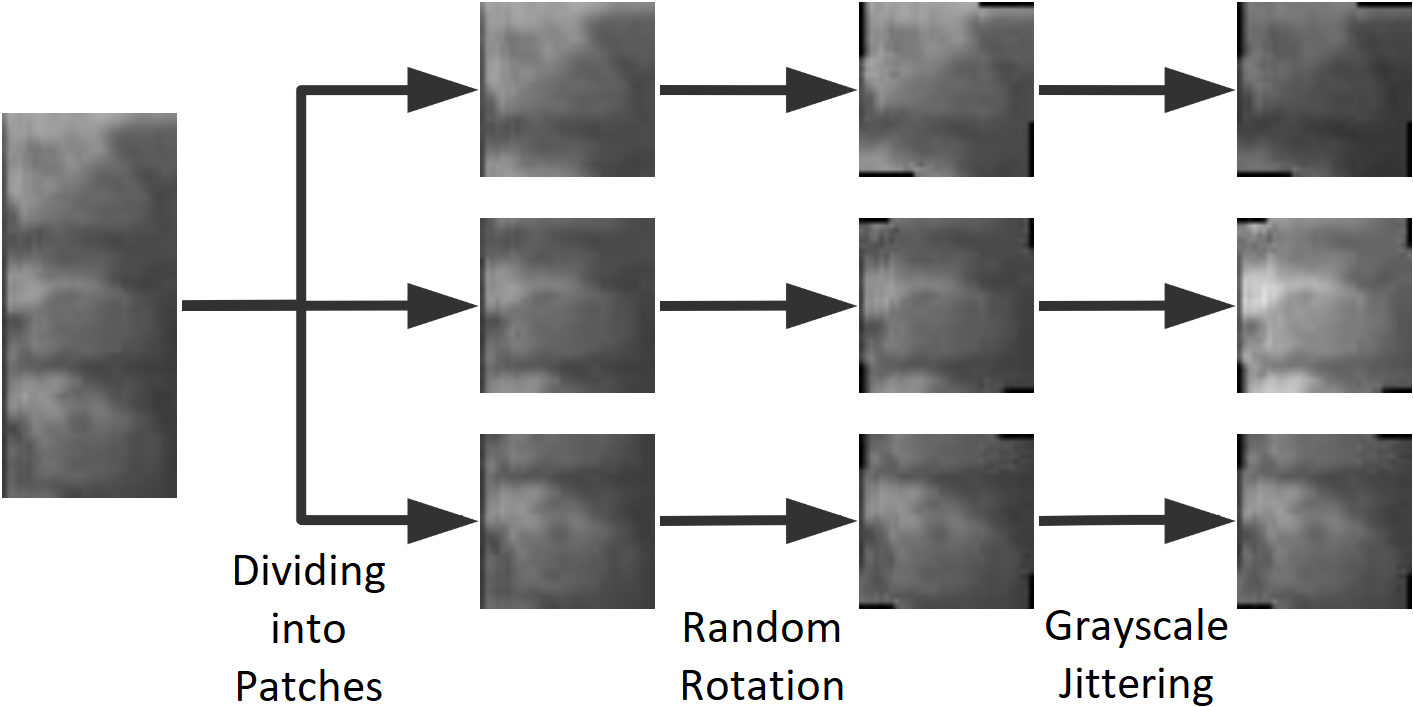}
    \caption{Patch augmentation on the training images.}
    \label{fig: patch_aug}
\end{figure}

After this step, we use image patches to create input image sequences for the TriPodNet model. Shapes of C2, C3 and C4 vertebrae determine the stage of the CVM. Therefore, we crop the segmented vertebrae image into $35\times35$ patches that contain only one vertebra as shown in Fig.~\ref{fig: model_diag}. Three patches are derived from each image before the patch augmentation. The location of each vertebra in the image is used to create the patches since C2, C3 and C4 are always in similar positions in any image. We used overlapped windows to create patches with the size of $35\times35$. Before using the patches in model training, we also benefit from patch augmentation where we rotate the images randomly by 5 degrees. The rotation is necessary to help the model generalize as the alignment of each vertebra depends on the posture of the patient. Moreover, we apply grayscale jittering on the patches. With the grayscale jittering, patches become brighter or darker, randomly. Patch augmentation is visualized in Fig.~\ref{fig: patch_aug}. Similar to other image classification methods, augmentation is only applied to the training images. We do not apply patch augmentation on testing images.

Next, instead of feeding the image patches directly to the ResNets, the edges of the vertebral body in the patches are emphasized using eight directional filters described in \cite{bagci2007low, bozkurt2014multi}. The outputs of directional filters are then fed to the ResNets. The same directional filters in \cite{atici2022fully} are used before ResNets to highlight the edges of CVM images. The motivation behind using the directional filters is to start the deep learning model using our prior domain knowledge. Since the angles of CVM bones are the distinguishing factor in estimating the CVM stages highlighting the edges of the bones in multiple directions will give an advantage to the deep network model. 

As pointed out above each pod of the TriPodNet is a ResNet-20. The structure of ResNet-20 with the directional filters and patch augmentation is summarized in Table~\ref{tab: resnet}. The ResNets are implemented in parallel. The output feature maps of ResNets are concatenated together with the age of the subject. We also use the chronological age information as the input to the fully connected layer. Obviously, the chronological age of a subject is correlated with the maturity of a patient. The age information is repeated six times in the vector and Gaussian noise with zero mean and 0.01 variance is added to secure its impact before the output layer. 

\begin{table}[htbp]
    \centering
    \begin{tabular}{ccc}
    \hline\noalign{\smallskip}
		\bf{Layer}&\bf{Output Shape}&\bf{Implementation Details}\\
         \noalign{\smallskip}\hline\noalign{\smallskip}
        PatchAug&$35\times35\times8$&-\\
        DirFilts&$35\times35\times8$&$7\times7,8$\\
		Conv1&$35\times35\times16$&$3\times3, 16$\\
		Conv2\_x&$35\times35\times16$&$\left[ \begin{array}{c} 3\times3, 16  \\ 3\times3, 16 \end{array}\right]\times 3$\\
		Conv3\_x&$17\times17\times32$&$\left[ \begin{array}{c} 3\times3, 32 \\ 3\times3, 32 \end{array}\right]\times 3$\\
		Conv4\_x&$8\times8\times64$&$\left[ \begin{array}{c} 3\times3, 64 \\ 3\times3, 64 \end{array}\right]\times 3$\\
		AAP&$64$&Adaptive Average Pooling\\
    \noalign{\smallskip}\hline\noalign{\smallskip}
	\end{tabular}
	\caption{Structure of each pod (ResNet with the directional filters) of the TriPodNet. PatchAug stands for patch augmentation. DirFilts stands for directional filters \cite{atici2022fully}.}
\label{tab: resnet}
\end{table}

% \begin{table}[htbp]
%     \centering
%     \begin{tabular}{ccc}
%     \hline\noalign{\smallskip}
% 		\bf{Layer}&\bf{Output Shape}&\bf{Implementation Details}\\
%          \noalign{\smallskip}\hline\noalign{\smallskip}
%         PatchAug&$35\times35\times8$&-\\
%         DirFilts&$35\times35\times8$&$7\times7,8$\\
% 		Conv1&$35\times35\times16$&$3\times3, 16$\\
% 		Conv1\_x&$35\times35\times16$&$\left[ \begin{array}{c} 3\times3, 16  \\ 3\times3, 16 \end{array}\right]\times 3$\\
% 		Res1\_x&$17\times17\times32$&$\left[\begin{array}{c}3\times3, 32\\3\times3, 32\end{array}\right] + [\begin{array}{c}1\times1, 32\end{array}]$\\
% 		Conv2\_x&$17\times17\times32$&$\left[ \begin{array}{c} 3\times3, 32 \\ 3\times3, 32 \end{array}\right]\times 2$\\
% 		Res2\_x&$8\times8\times64$&$\left[\begin{array}{c}3\times3, 64\\3\times3, 64\end{array}\right] + [\begin{array}{c}1\times1, 64\end{array}]$\\
% 		Conv3\_x&$8\times8\times64$&$\left[ \begin{array}{c} 3\times3, 64 \\ 3\times3, 64 \end{array}\right]\times 2$\\
% 		AAP&$64$&Adaptive Average Pooling\\
%     \noalign{\smallskip}\hline\noalign{\smallskip}
% 	\end{tabular}
% 	\caption{Structure of each pod (ResNet with the directional filters) of the TriPodNet. PatchAug stands for patch augmentation. DirFilts stands for directional filters \cite{atici2022fully}.}
% \label{tab: resnet}
% \end{table}

We selected the TriPodNet with random translation and AutoContrast augmentation methods \cite{pytorchautocontract} as the best network model in our dataset based on our experiments. In the next section, we present our experimental results. 

\section{Experimental Results}

We studied different networks, and different data augmentation methods to determine the best possible network structure.  As introduced in \cite{pan2022multipod}, MultiPodNet can be constructed from two or more pod networks processing the input in parallel. Outputs of pod networks can be combined by adding the feature maps or by concatenation that produced a higher accuracy than adding the feature maps of individual networks.
%Furthermore, different image patching strategies can be studied. 
We studied a single channel ResNet, DuPodNet which consists of two pod networks, TriPodNet, and QuadPodNet which consists of four parallel pod networks. The TriPodNet with input overlapping input image patching strategy as shown in Fig.~\ref{fig: model_diag} produced the highest accuracy in our dataset.
%and StackNet for comparison.
%DuPodNet uses two ResNet pods in the main part of the model, and QuadPodNet uses four. 
%StackNet feeds all three patches stacked to three pods. To compare and prove the result, 
We compare the TripodNet constructed from ResNet-20s~\cite{he2016deep} with the Swin Transformer~\cite{liu2021swin}, Xception~\cite{chollet2017xception}, MobileNet-V1~\cite{howard2017mobilenets}, MobileNet-V2~\cite{sandler2018mobilenetv2}, and the custom designed CNN with the directional filters that achieved the previous best result in~\cite{atici2022fully}. 
%The last variable is the data augmentation techniques. ConvNext~\cite{liu2022convnet}, the paper inspired \cite{pan2022multipod}, mentions more advanced augmentation techniques can increase the model performance on unseen data. 
In addition to random translation and AutoContrast data augmentation methods we studied "randAugment", "AugMix" methods and without any augmentation. 
%Notice that patch augmentation is not affected by the choice of data augmentation method.

To train the networks, an SGD optimizer with a weight decay of 0.0001 and momentum of 0.9 is used. These models are trained with a batch size of 32, an initial learning rate of 0.1 for 100 epochs, and the learning rate is reduced to 1/10 at epochs 25, 50, and 75. The experiments are implemented using PyTorch in Python 3.

%We first investigate the improvement with directional filters and different augmentation methods for the TripodNet.
%Then, we will compare the TripodNet with DuPodNet, StackNet, and QuadPodNet using the best pre-processing data augmentation methods to show that the three-pod structure is the optimal structure for the CVM classification task. Finally, we will compare the TripodNet with other state-of-art networks. 

The accuracy results of the TripodNet with and without directional filters and with different augmentation methods are summarized in Table~\ref{tab: Compare}. Directional filters improve TripodNet's accuracy by 4.88\% (from 76.29\% to 81.17\%) in the dataset containing female subjects. Similarly, they improve the accuracy by 4.17\% (from 71.15\% to 75.32\%) in the dataset containing male subject images. As pointed out above we augment both the entire image and the patches.
We trained the network using random translations with AutoContrast as the data augmentation method as shown in Table~\ref{tab: Compare2}. Furthermore, we augment the input image patches to make the system robust to changes in posture and exposure as shown in Table~\ref{tab: Compare} during training.
%shows that with the directional filters, data augmentation, and patch augmentation, the TripodNet reaches the highest accuracy on both female and male datasets. For example, in the cases with data augmentation and patch augmentation, directional filters improves TripodNet's female accuracy by 4.88\% (from 76.29\% to 81.17\%) and male accuracy by 4.17\% (from 71.15\% to 75.32\%).
\begin{table}[htbp]
    \centering
    \begin{tabular}{ccccc}
    \hline\noalign{\smallskip}
        \bf{Directional} & \multicolumn{2}{c}{\bf{Augmentation}} & \multicolumn{2}{c}{\bf{Accuracy}} \\
         \bf{Filters} & \bf{Data}  & \bf{Patch} & \bf{Female} & \bf{Male} \\
     \noalign{\smallskip}\hline\noalign{\smallskip}
$\times   $&$\times   $&$\times   $& 67.05\%    & 65.38\%  \\
$\times   $&\checkmark&$\times   $& 75.11\%    & 70.19\%  \\
\checkmark &\checkmark&$\times   $& 78.64\%    & 70.19\%  \\
$\times   $&\checkmark&\checkmark & 76.29\%    & 71.15\%  \\
\checkmark &\checkmark&\checkmark & \textbf{81.17}\%    & \textbf{75.32}\%  \\
\noalign{\smallskip}\hline\noalign{\smallskip}
	\end{tabular}
\caption{Accuracy results of the TriPodNet with and without directional filters, augmentation of entire input images, and augmentation of image patches.}
\label{tab: Compare}
\end{table}

\begin{table}[htbp]
\centering
    \begin{tabular}{ccc}
    \hline\noalign{\smallskip}
        \bf{Data Augmentation Method}  & \bf{Female} & \bf{Male} \\
     \noalign{\smallskip}\hline\noalign{\smallskip}
No augmentation   & 68.23\%    & 66.34\%  \\
Random augmentation   & 74.11\%    & 70.18\%  \\
AugMix  & 77.64\%    & 69.23\%  \\
\bf{Random translation, AutoContrast}  & \bf{81.17}\%    & \bf{75.32}\%  \\
\noalign{\smallskip}\hline\noalign{\smallskip}
	\end{tabular}
\caption{Accuracy results of various data augmentation methods. Directional filters and patch augmentation are applied. AugMix, AutoContrast are PyTorch commands \cite{pytorchautocontract}.}
\label{tab: Compare2}
\end{table}

The accuracy of different MultiPod networks is presented in Table~\ref{tab: Compare3}. TriPodNet achieves the highest accuracy. QuadPodNet has more parameters compared to TriPodNet but it produced a lower accuracy. This may be due to the small dataset size and training issues.
%and adding more pods reduces the performance.
StackNet in Table~\ref{tab: Compare3} has also three parallel pod networks. 
In the StackNet, we stack the input patches and feed the augmented stack patches to the pod networks at the same time. 
\begin{table}[htbp]
    \centering
    \begin{tabular}{cccc}
    \hline\noalign{\smallskip}
        \bf{Model}& \bf{Parameters}  & \bf{Female} & \bf{Male} \\
     \noalign{\smallskip}\hline\noalign{\smallskip}
ResNet  & 0.27M & 75.29\%    & 73.07\%  \\
DuPodNet  & 0.54M & 75.29\%    & 68.27\%  \\
StackNet   & 0.82M & 80.03\%    & 70.19\%  \\
\textbf{TriPodNet}   & 0.81M & \bf{81.17}\%    & \bf{75.32}\%  \\
QuadPodNet   & 1.08M & 76.84\%    & 71.43\%  \\
\noalign{\smallskip}\hline\noalign{\smallskip}
	\end{tabular}
\caption{Accuracy results of MultiPod networks. DuPodNet uses two ResNet pods, TriPodNet uses three, and QuadPodNet uses four. StackNet feeds all three patches stacked to three pods.}
\label{tab: Compare3}
\end{table}

\begin{figure}[htbp]
\centering
\subfloat[\label{fig: acc_comp_female} Female patients.]{\includegraphics[width=0.5\linewidth]{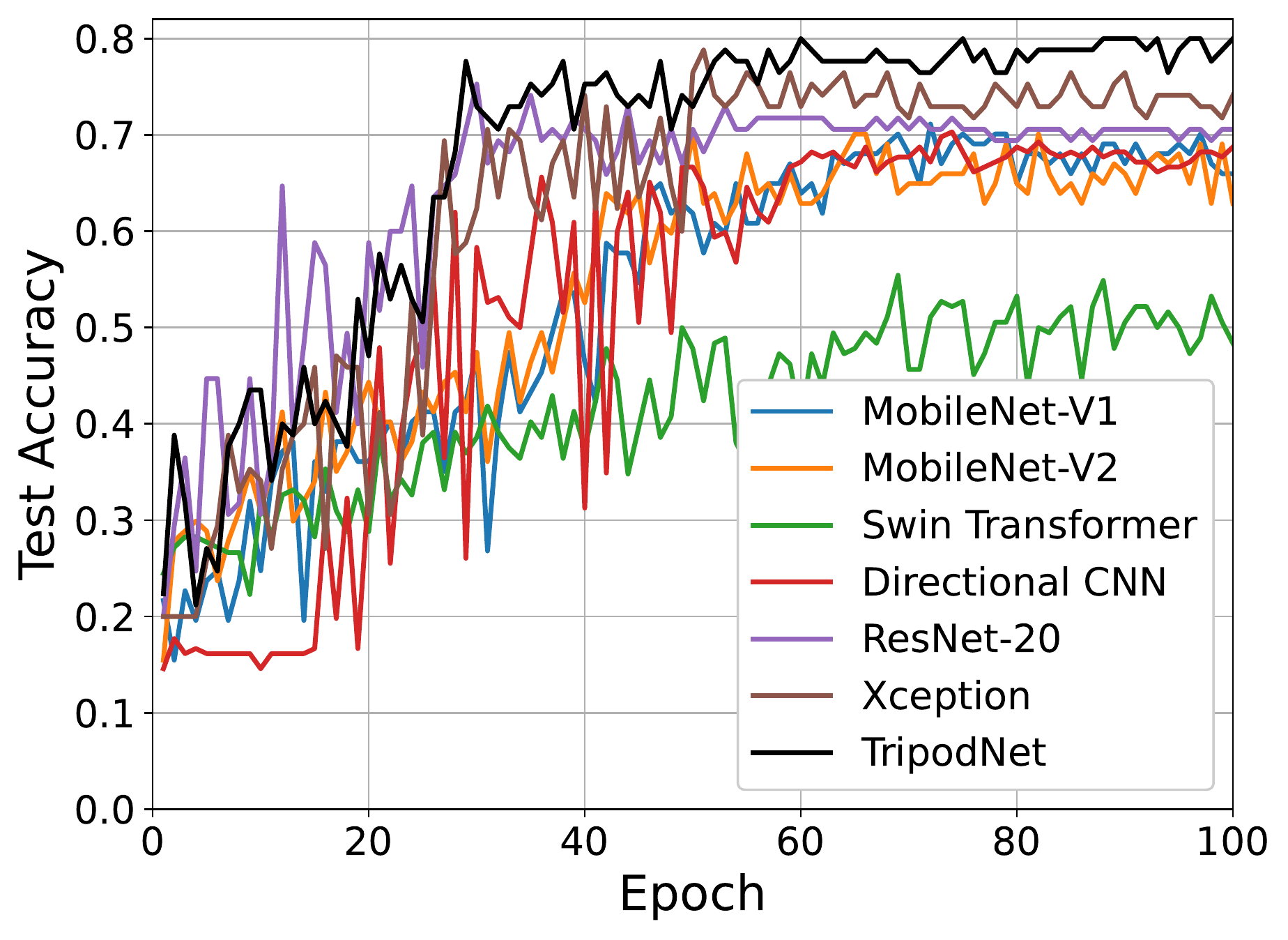}}
\subfloat[\label{fig: acc_comp_male}Male patients.]{\includegraphics[width=0.5\linewidth]{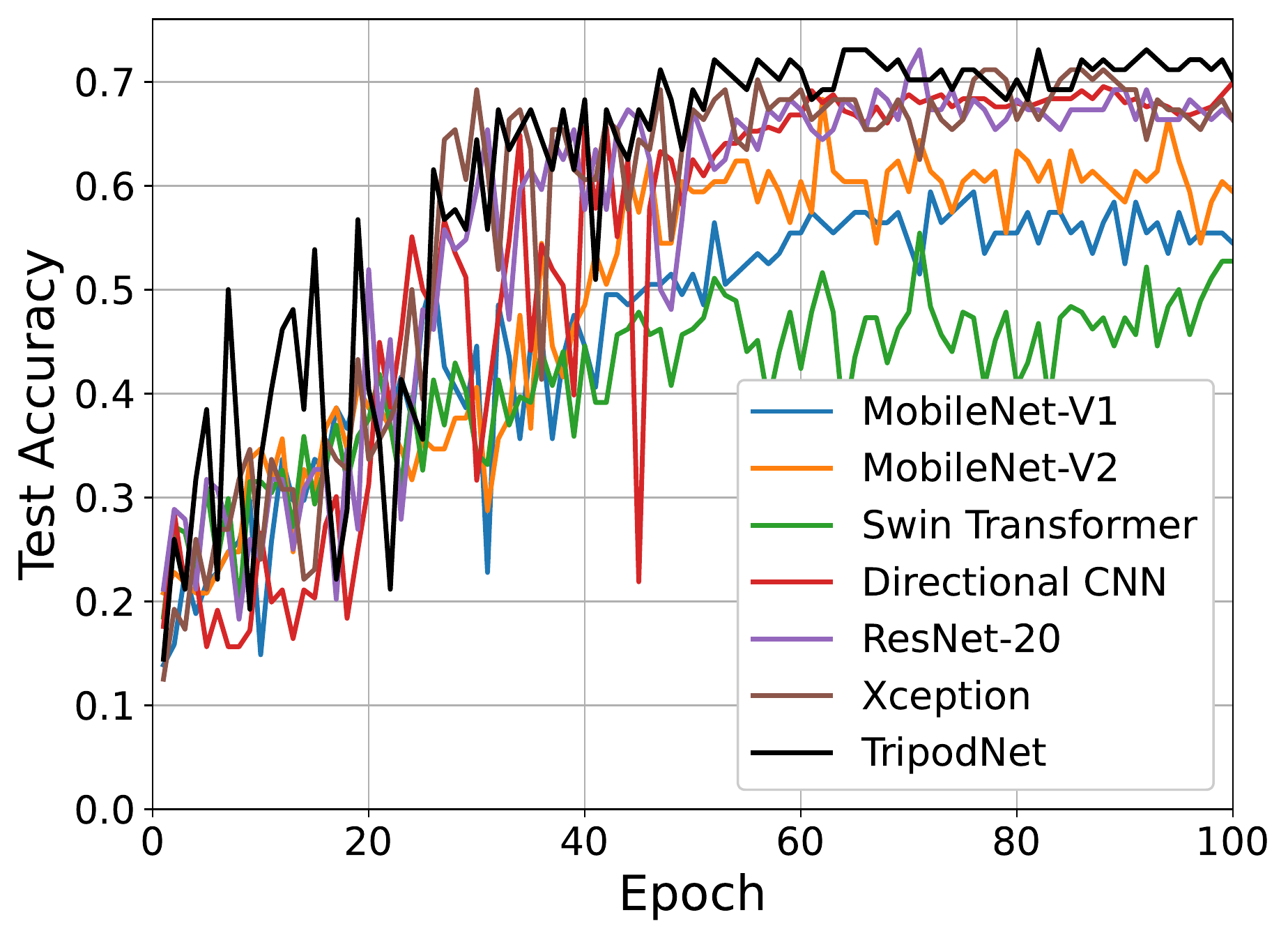}} 

\caption{Test accuracy versus the number of epochs of different networks. The TripodNet reaches the highest accuracy on both genders.}
\label{fig: acc_comp}
\end{figure}

Comparison of the TripodNet with other state-of-art networks is presented in Table~\ref{tab: Compare4} and Fig.~\ref{fig: acc_comp}. Fig.~\ref{fig: acc_comp} shows the test accuracy versus the number of epochs of different networks. In Table~\ref{tab: Compare4}, we compare the TripodNet with ResNet-20~\cite{he2016deep}, Swin Transformer~\cite{liu2021swin}, Xception~\cite{chollet2017xception}, MobileNet-V1~\cite{howard2017mobilenets}, MobileNet-V2~\cite{sandler2018mobilenetv2}, and the custom-designed CNN with the directional filters~\cite{atici2022fully}. We adjust the image sizes to pre-trained networks: we remove some of the downsampling layers in the MobileNet-V1, MobileNet-V2, and Xception. We implement the Swin Transformer using two stages, where each stage contains a patch merging layer with 4 Swin Transformer blocks. Moreover, it is well-known that after pre-training on a large dataset, a transformer can get a much higher accuracy (this is known as transfer learning). Therefore, to compare with the Swin Transformer, we also use the ImageNet-1k-pre-trained Swin-s and Swin-t transformers. In these cases, the input tensors are interpolated to $224\times224$ to adjust the CVM image sizes to transformer input sizes. Table~\ref{tab: Compare4} and Fig.~\ref{fig: acc_comp} show that our proposed TripodNet produces the highest accuracy in our data set. The TripodNet gets better results than the pre-trained Swin-s and Swin-t transformers. This may be due to the "tiny" size of the CVM data set. It is probably too small for the transformer networks. Another reason is that transfer learning may not work properly because the CVM images are quite different from the ImageNet images.

\begin{table}[htbp]
\centering
\begin{tabular}{cccc}
\hline\noalign{\smallskip}
\bf{Model} & \bf{Parameters}  & \bf{Female} & \bf{Male} \\
\noalign{\smallskip}\hline\noalign{\smallskip}
%MobileNet           & 2.25M & 62.35\%    & ?? \%  \\
MobileNet-V1           & 3.21M & 73.20\%    & 60.40\%  \\
MobileNet-V2           & 2.29M & 72.16\%    & 68.32\%  \\
Swin Transformer    & 4.12M & 63.58\%    & 62.50\%  \\
Pre-trained Swin-s  & 48.8M & 75.26\%    & 73.27\%  \\
%Swin-s & 48.8M & 60.82\%    & 70.30\%  \\
Pre-trained Swin-t & 28.5M & 72.16\%    & 74.26\%  \\

%Swin-t & 28.5M & 65.98\%    & 73.27\%  \\
Directional CNN~\cite{atici2022fully}         & 0.71M & 70.33\%    & 71.88\%  \\
ResNet-20              & 0.27M & 75.29\%    & 73.07\%  \\
Xception            & 33.0M & 78.82\%    & 71.15\%  \\
%Xception            & 9.52M & 72.16\%    & 68.32\%  \\
%Xception            & 20.82M & 62.89\%    & 69.31\%  \\
\bf{TripodNet}          & 0.81M & \bf{81.17}\%    & \bf{75.32}\%  \\
\noalign{\smallskip}\hline\noalign{\smallskip}
\end{tabular}
\caption{Accuracy results of various networks. Swin-s and Swin-t were pre-trained on ImageNet-1K.} 
\label{tab: Compare4}
\end{table}

\begin{figure}[htbp]
\centering
\subfloat[\label{fig: conf_female}Female patients.]{\includegraphics[width=0.5\linewidth]{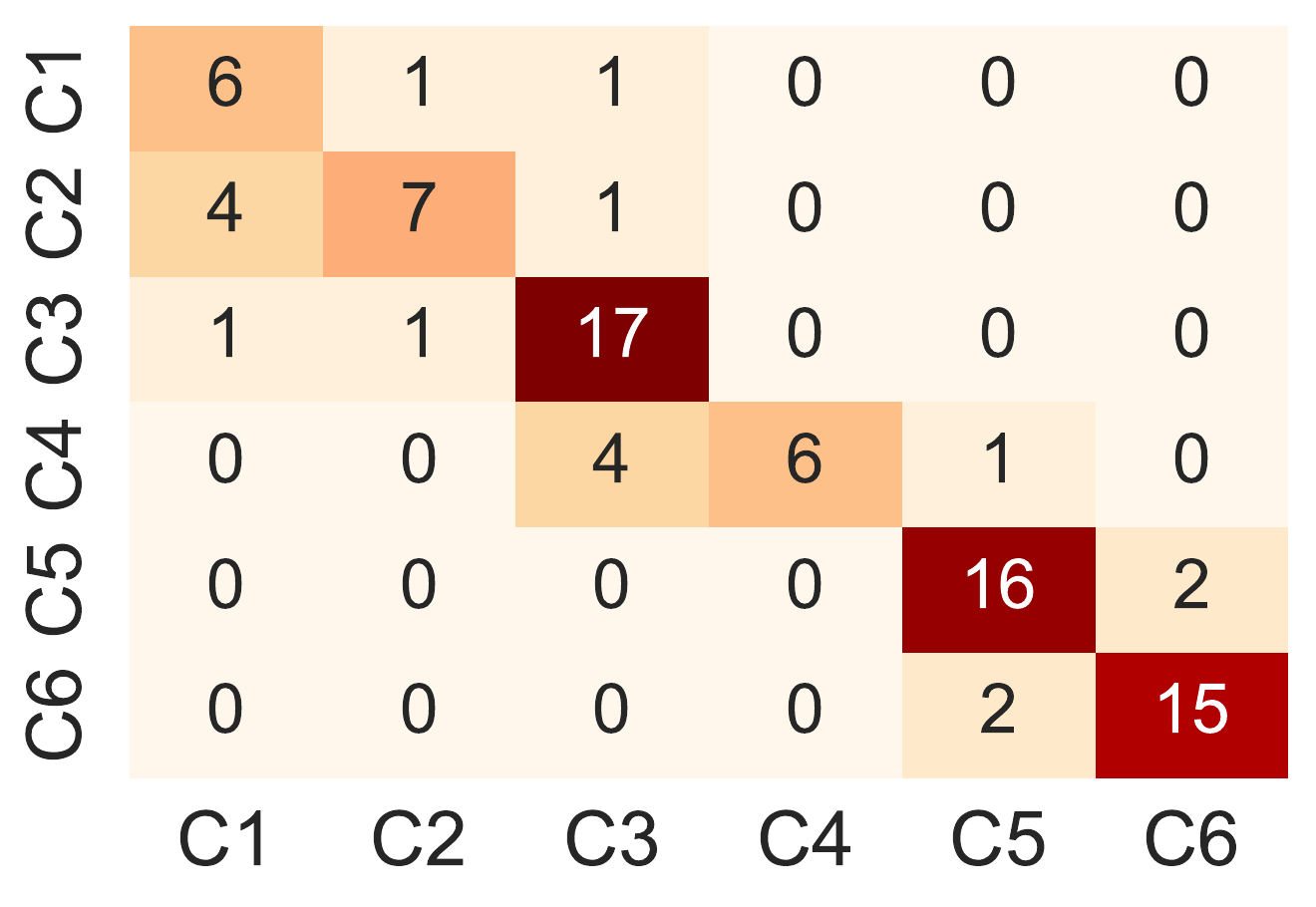}}
\subfloat[\label{fig: conf_male}Male patients.]{\includegraphics[width=0.5\linewidth]{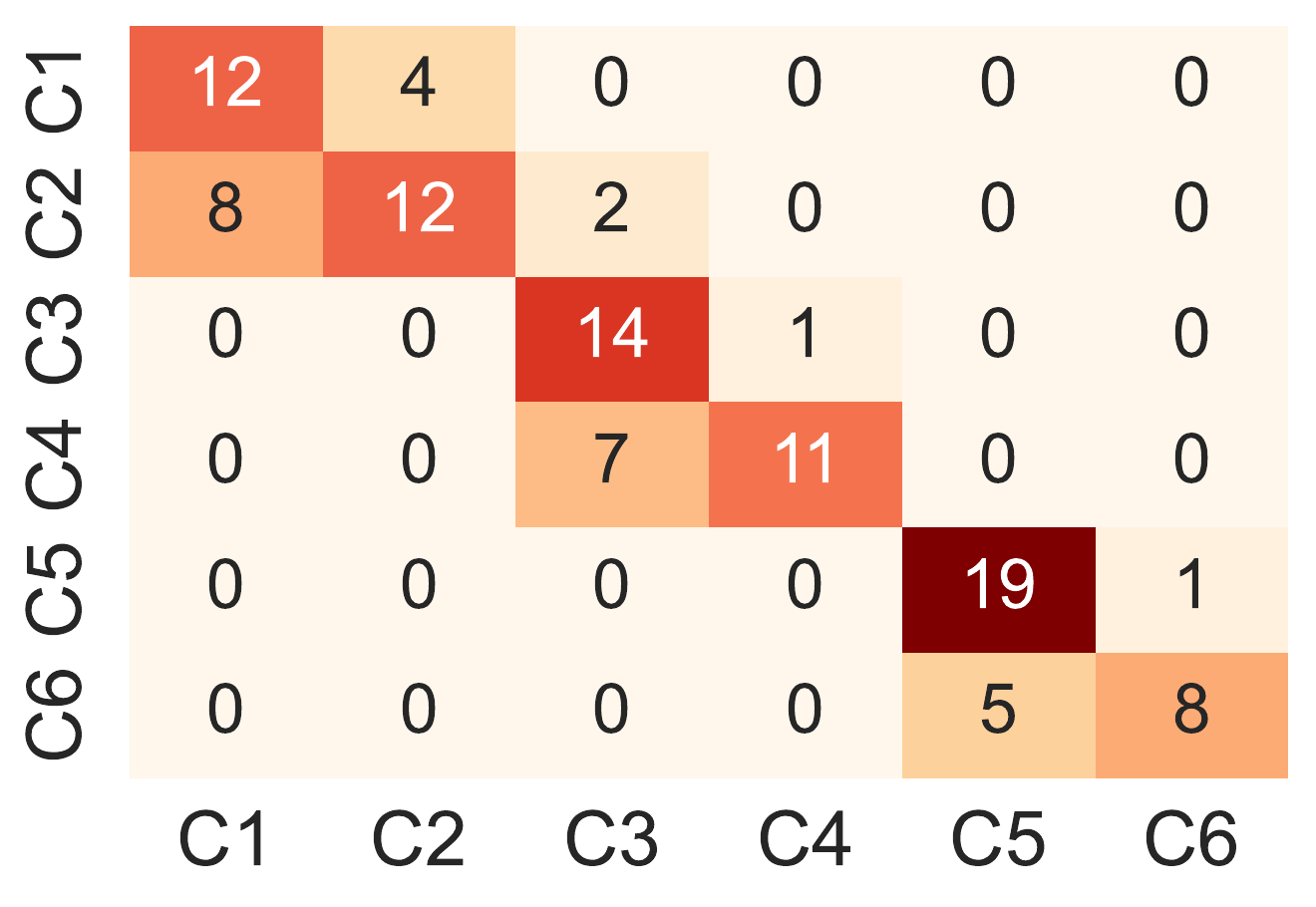}}
\caption{The confusion matrices of the TriPodNet.}
\label{fig: confusion}
\end{figure}

Fig.~\ref{fig: confusion} shows the confusion matrices of the TriPodNet in the male and female patient datasets.

\section{Conclusion}
In this paper, we present a new method for CVM classification. We introduce a novel neural network which is a combination of three parallel networks for this purpose. Similar to the transformer networks, TriPodNet performs its computations in parallel using traditional CNNs and the output feature maps of parallel CNNs are combined using a fully connected layer to produce the final result. We also compared the results of two, three, four, or more parallel networks. The TriPodNet with three parallel ResNet-20s produced the best accuracy result.

The transformer networks did not produce as good results as the TriPodNet. This is probably because our tiny data set is too small for the transformer networks. 
\ninept

\bibliographystyle{IEEEbib}
\bibliography{main}

\end{document}